# Continuous tuning & thermally induced frequency drift stabilisation of time delay oscillators such as the optoelectronic oscillator

Mehedi Hasan[1]*, Charles Nicholls[2], Keegan Pitre[2], Trevor Hall[1]

[1]Photonic Technology Laboratory, University of Ottawa, 25 Templeton Street, Ottawa, ON, Canada, K1N 6X1

[2]NANOWAVE Technologies Inc., 6 Gurdwara Rd, Nepean, ON, Canada, K2E 8A3

*mhasa067@uottawa.ca

## Abstract

Delay line oscillators based on photonic components, such as the optoelectronic oscillator (OEO), offer the potential for realization of phase noise levels orders of magnitude lower than achievable by conventional microwave sources. Fibre optic-based delay lines can realize the large delay required for low phase noise systems whilst simultaneously achieving insertion loss levels that can be compensated by available microwave and photonic amplification technologies. However, the long fibre is vulnerable to environmental perturbations such as mechanical vibrations and variations in ambient temperature, which result in short term fluctuations and thermally induced drift of the oscillation frequency. The phase shifter used conventionally to adjust the frequency of an OEO to enable phase lock has a finite range that is insufficient to compensate the delay change resulting from operational temperature ranges. A solution to continuous tuning without mode-hopping and to compensation of thermally induced frequency drift without loss of lock of a time delay oscillator is proposed. The basic concept is to introduce a tuning mechanism that works with Cartesian co-ordinates on the complex plane and to avoid explicit use of polar co-ordinates. Consequently, the transmission of the tuning component may traverse the unit circle in either direction multiple times without range limitation to the phase. Thereby tuning by mode-hopping is avoided and expedients to stabilisation, such as the use of tunable lasers within a control loop or precision temperature stabilisation measures are not required. The concept is verified by Simulink simulations. The method has been experimentally tested successfully using a prototype OEO phase locked to a system reference. Solid lock was maintained even when the OEO was placed in an oven and cycled over a temperature range from ambient to 80 °C.

## 1. Introduction

The phase noise and frequency stability of the system oscillator / clock is of paramount importance in applications such as optical and wireless communications, high-speed digital electronics, RADAR, and astronomy. Delay line oscillators based on photonic components, of which the optoelectronic oscillator (OEO) [1] is the most suited to practical deployment, offer the potential for realization of phase noise levels orders of magnitude lower than achievable by conventional microwave sources. The OEO generates microwave carriers using an RF photonic link consisting of laser; optical intensity modulator; optical fibre delay line; photo-receiver; RF amplifier and bandpass filter, which drives the modulator; closing the loop and sustaining oscillation (see Figure 1). Fibre optic-based delay lines can realize the large delay required for low phase noise systems whilst simultaneously achieving insertion loss levels that can be compensated by available microwave and photonic amplification technologies. However, the long fibre is vulnerable to environmental perturbations such as mechanical vibrations and variations in ambient temperature, which result in short term fluctuations and thermally induced drift of the oscillation frequency.

Approaches to improved long-term stability include acoustic isolation [2]; thermal stabilisation [3]; temperature insensitive speciality fibre [4,5]; tuneable laser within a control loop [6]; frequency division and multiplication [7] and phase locking to a system reference [8,9]. None of these expedients alone is a complete solution and all but the phase locked loop (PLL) add significant complexity, size, weight, power consumption, and cost. Whereas an oven-controlled quartz crystal oscillator (OCXO), even when multiplied to microwave frequencies, offers superior phase noise performance at frequencies within a small neighbourhood of the carrier, an OEO offers superior performance at frequencies outside this neighborhood. A state-of-the art 100 MHz OCXO [10] achieves a phase noise of -177 dBc/Hz at 10 kHz which increases to -137 dBc/Hz at 10 kHz when multiplied to 10 GHz. So, the very best in-class OCXO is outperformed by a 10 GHz OEO and at 20 GHz the phase noise advantage is increased by 6 dB as the OEO does not suffer multiplicative increase in phase noise leading to increasing performance benefit with increasing frequency [1].

It is a requirement of systems to phase lock all internal sources to a common reference to ensure coherent operation and provide long-term stability. Engineering the PLL to take advantage of the superior phase noise of a free OEO at higher offset frequencies requires the control bandwidths to be necessarily small to prevent the source phase noise being degraded by the multiplied phase noise of the reference oscillator. However, the phase shift used to tune an OEO within a PLL is necessarily bounded and typically insufficient to compensate thermally induced frequency drift caused by operational temperature ranges; without additional thermal control, a PLL may lose lock within a few degrees of temperature variation.

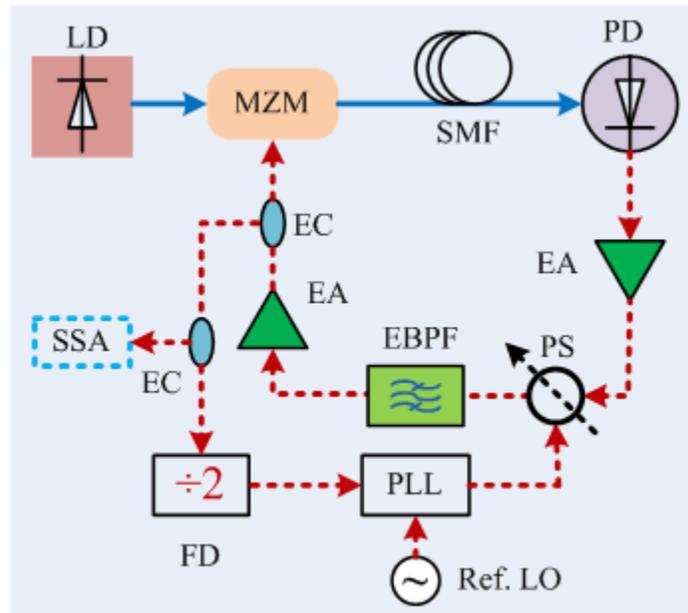

*Figure 1 Schematic diagram of a phase-locked optoelectronic oscillator; LD laser diode, MZM Mach-Zehnder modulator, SMF single mode fibre delay line, PD photodetector; EA electrical amplifier; PS phase shifter, EBPF electrical bandpass filter, EC electrical coupler; PLL phase lock loop; FD, Frequency divider; SSA, signal source analyzer; Ref. LO, reference local oscillator.*

This paper describes a solution to continuous tuning without mode-hopping and to compensation of thermally induced delay drift without loss of lock of a time delay oscillator (TDO) of which the OEO is an example. The innovation is to introduce components that work with Cartesian co-ordinates $(x, y)$ on the complex plane and to avoid explicit use of polar co-ordinates $(\rho, \theta)$. Consequently, the transmission of the tuning component may traverse the unit circle in either direction multiple times without range limitation to the phase or any requirement to unwrap its principal part. The concept has the merit that tuning by mode-hopping is avoided and expedients to stabilisation, such as the use of tunable lasers within a control loop or precision temperature stabilisation measures are not required.

The paper is organized as follows. In Section 2 it is explained how continuous tuning without mode hopping of a TDO may be achieved; providing a prelude to the consideration in Section 3 of the phase locking to a system reference of a TDO subject to substantial delay drift without loss of lock. Section 4 presents the results of simulations provided by Simulink models of the TDO and PLL which conclusively support the analysis of Section 2 and Section 3. Finally, Section 4 concludes the paper with a summary and discussion of the main findings.

## 2. Continuous tuning

The phase $\phi$ of the oscillation circulating within a time delay oscillator satisfies the Barkhausen round-trip phase condition:

$$\phi(t) = \theta + \phi(t - \tau)$$

*Equation 1*

where $\tau$ is the delay contributed by the optical fibre delay line and $\theta$ is an additional phase contributed by the RF components within the path, which includes a controlled phase-shifter.

For a single mode oscillation of frequency $\omega$, the phase has the form:

$$\phi(t) = \omega t$$

*Equation 2*

Substitution of Equation 2 into Equation 1 yields an equation that determines the frequency $\omega$ in terms of the delay $\tau$ and the tuning phase shift $\theta$:

$$\omega \tau - \theta = 2n\pi \quad ; \quad n \in \mathbb{Z}$$

*Equation 3*

The integer $n$ indexes the multitude of possible oscillation modes. An RF filter within the loop may admit several modes within its passband but the mode closest in frequency to the resonant frequency of the RF-selection filter is favoured in a winner-takes-all competition for the reservoir of energy provided by the sustaining amplifier which determines the specific single oscillating mode.

It is clear from the Barkhausen condition that for a constant additional phase[1] adjacent modes are spaced in frequency by:

$$\Delta f = 1/\tau$$

*Equation 4*

and conversely a phase change of:

$$\Delta \theta = 2\pi$$

*Equation 5*

will tune the frequency of the modes by one free-spectral range (FSR) $\Delta f$, i.e., the comb of oscillation frequencies is translated by an adjustment of the tuning phase and the comb returns to its original alignment after a $2\pi$ tuning phase change.

An adiabatic adjustment of the phase shift may therefore be used to fine-tune the oscillator. However, if the frequency of oscillation is pulled too far from the passband centre of the RF filter, there is a risk the oscillator will mode-hop, i.e., the oscillation in the pulled mode will cease and a mode better favoured by the RF filter will assume oscillation. Continuous tuning without mode hops requires ganged tuning of the RF filter and the controlled phase-shift to maintain alignment of the frequency of the oscillating mode and the resonant frequency of the RF selection filter. *Prima facie, continuous tuning over an indefinite range requires a controlled phase-shift that has an indefinite range*. The controlled phase-shift range in practice is limited but, if the range exceeds $2\pi$, a tuneable RF filter can provide coarse tuning via mode-hops while the phase-shift element can provide fine tuning over one FSR. However, tuning by discontinuous mode hops is not a satisfactory solution for some applications.

---

[1] In practice the RF components and in particular the RF selection filter have a dispersive frequency response. The additional phase is consequently weakly dependent on the oscillation frequency. The condition that the additional phase is held constant implies a small adjustment of the controlled phase-shift.

The Barkhausen phase (and amplitude) condition for oscillation arise from the complex condition:

$$\exp[-i(\omega\tau - \theta)] = \exp(i\theta)\exp(-i\omega\tau) = 1$$

*Equation 6*

and this may be written:

$$z \exp(-i\omega\tau) = 1$$

*Equation 7*

where:

$$z = x + iy \quad ; \quad x = \cos(\theta) \quad , \quad y = \sin(\theta)$$

*Equation 8*

Noting:

$$\Re\{z \exp(-i\omega\tau)\} = x \cos(\omega t) + y \sin(\omega t)$$

*Equation 9*

it is observed that the phase-shift corresponding to the control variable $z$ may be applied to the RF oscillation via an IQ modulator[2] by applying $x$ to the in-phase (I) port and $y$ to the quadrature-phase (Q) port. Given the narrowband oscillation, the quadrature component of the RF oscillation may be derived from the in-phase component of the RF oscillation by a quarter wave delay or a 90° hybrid or some other equivalent means[3].

An arbitrarily large variation of $\theta$ is mapped into a strictly bounded variation of the components $x$ and $y$ as $z$ ranges over the unit circle. While it remains the case that the fine tuning achieved remains limited to an FSR as $z$ ranges over the unit circle, the frequency of oscillation is a *continuous* function of $z$. Traversing the unit circle in the anti-clockwise (clockwise) direction increases (decreases) the frequency of the oscillating mode. One complete circuit increases (decreases) the frequency of oscillation by one FSR. The oscillating mode at $\theta = 2\pi$ is then indistinguishable from the adjacent mode to the oscillating mode at $\theta = 0$ because the number of RF cycles within the delay line has been incremented (decremented) by one due to the increase (decrease) in oscillation frequency by one FSR. Provided the alignment in frequency of the resonant peak of the RF filter with the frequency of oscillation is broadly maintained and the tuning rate is sufficiency slow on the scale of the round-trip time, the winner-takes-all competition ensures that the frequency of oscillation will continue to increase (decrease) as the control continues its anti-clockwise (clockwise) trajectory. Thereby continuous tuning without mode-hops is achieved.

## 3. Thermally induced frequency drift stabilisation

Attention is now turned to a related situation in which the frequency of oscillation is locked to a system reference using a phase-locked loop (PLL) but the delay $\tau$ is subject to fluctuations and thermally induced drift. The orthodox approach uses the phase $\theta$ as the control variable. Since in a narrow band system it is only necessary to compensate modulo $2\pi$ the phase drift contributed by the change in delay and provided the tuning phase shifter has a range greater than $2\pi$, then the thermally induced drift can be stabilised in principle[4]. However, an instantaneous fly-back by $2\pi$ is necessary as the controlled phase approaches a limit of its range. In practice, a fly-back transient disrupts the phase lock and in the presence of fluctuations multiple transients could occur in rapid sequence. In theory, a type II PLL has an unbounded locking range but to realise that potential the oscillator must possess an unbounded tuning range. If there is no fly-back and the drift has the same sign for a sufficient interval of time, the controller or tuning phase-shifter will reach a limit of their range and lock will lost. Moreover, the PLL will not lock at start up if the oscillator is detuned from the reference by more than the tuning range (~ 1 FSR) necessitating a search algorithm. In common with the continuous tuning problem, the difficulties originate from the branch cut inherent in the use of a polar co-ordinate system on the complex plane and the solution is to use a controller and a vector modulator that both work with a Cartesian co-ordinate system $(x, y)$ in the complex plane.

---

[2] An IQ modulator is known as a vector modulator by the microwave technology community.
[3] If the oscillator is tuneable over a broad band, the means of deriving the quadrature component may need to be adjustable to provide $\pi/2$ phase-shift at the frequency of oscillation.
[4] The electronic phase shifter device used in the prototype OEO provides a phase range of $0° - 410°$.

To provide intuition, consider a type-II phase locked loop (PLL), which has a loop filter containing an integrator with real valued output $\theta$ given by:

$$\theta(t) = \theta(0) + \frac{1}{\tau}\int_0^t v(s)ds$$

*Equation 10*

where $v$ is the input and $\tau$ is a characteristic time constant. Set:

$$z = \rho \exp(i\theta)$$

*Equation 11*

then:

$$\tau \frac{dz}{dt} = ivz$$

*Equation 12*

The quantity $\rho = |z|$ is conserved by a solution of Equation 12. In principle, provided $z(t)$ commences on the unit circle, it will remain on the unit circle for all time. However, it is prudent to add a magnitude restoration mechanism that ensures that the unit circle is an attractor so that any perturbation of $z(t)$ away from the unit circle due to noise will decay with time. A suitably modified equation is given by:

$$\tau \frac{dz}{dt} = \mu(1 - |z|^2)z + ivz$$

*Equation 13*

where $\mu$ controls the strength of the magnitude restoration process. Equation 13 is a special case of an equation describing the Stuart-Landau oscillator.

Substitution of Equation 11 into Equation 13 and equating real and imaginary parts separates Equation 13 into two independent real equations:

$$\tau \frac{d\theta}{dt} = v$$

*Equation 14*

$$\tau \frac{d\rho}{dt} = \mu(1 - \rho^2)\rho$$

*Equation 15*

Equation 10 provides the general solution to Equation 14. Consequently, the evolution of the phase is unaltered by the magnitude restoration process. The evolution of $\rho$ may be found by recasting Equation 15 into an equivalent but linear equation in $1/\rho^2$ given by:

$$\frac{\tau}{2\mu}\frac{d}{dt}(1/\rho^2) + (1/\rho^2) = 1$$

*Equation 16*

which has general solution:

$$\frac{1}{\rho^2(t)} = 1 + \left(\frac{1}{\rho^2(0)} - 1\right)\exp(-2\mu\, t/\tau)$$

*Equation 17*

Consequently $\rho^2(t) \rightarrow 1$ monotonically as $t \rightarrow \infty$ for all $\rho(0) \neq 0$ so the unit circle is an attractor[5].

A Stuart-Landau oscillator may replace the integrator of a type-II PLL while providing the in-phase and quadrature phase controls $(x, y)$ to a vector modulator for continuous tuning, thus restoring the essentially[6] unbounded locking range of a type II PLL. The instantaneous frequency of the Stuart-Landau oscillator $\nu/\tau$ is set by the output of the phase-sensitive detector via a loop amplifier & filter designed to ensure the stability of the PLL. *Negative and positive frequencies are distinguished as clockwise and anti-clockwise rotation respectively.* The vector modulator within the oscillator acts as a mixer that outputs a sideband at a frequency $\omega + \nu/\tau$ while suppressing the carrier at frequency $\omega$ and the sideband at frequency $\omega - \nu/\tau$. The oscillator accumulates the frequency shift $\nu/\tau$ every round trip which results in a ramp of its oscillation frequency. Consequently, the combined Stuart-Landau oscillator and vector modulator is equivalent to an ideal (infinite range) linear phase integrator and phase-shifter. Whereas the use within a PLL of a combination of a linear integrator and phase shifter will result in the loss of lock at a limit of the phase shifter range; the combined Stuart-Landau oscillator and vector modulator has bounded controls. Consequently, the PLL may track a chirped frequency source with an instantaneous frequency confined within the passband of the RF filter or compensate for arbitrarily large delay drift provided the chirp or drift is slowly varying on the scale of the round-trip time of the oscillator. A Stuart-Landau oscillator is referred to alternatively as a Stuart-Landau integrator in this work whenever it is more appropriate to its function and to avoid confusion with the controlled time delay oscillator.

## 4. Simulation

The standard approach to modelling phase locked loops is to use the phase as the state-variable. Typically, a controlled single-mode oscillator is assumed that may be modelled as a simple phase integrator. However, a TDO is fundamentally a multimode oscillator. A Leeson model of a TDO succeeds in capturing the multimode dynamics of a TDO whilst retaining phase as the state variable. Consequently, phase models are useful for time domain studies of the dynamical behaviour of a PLL-OEO.

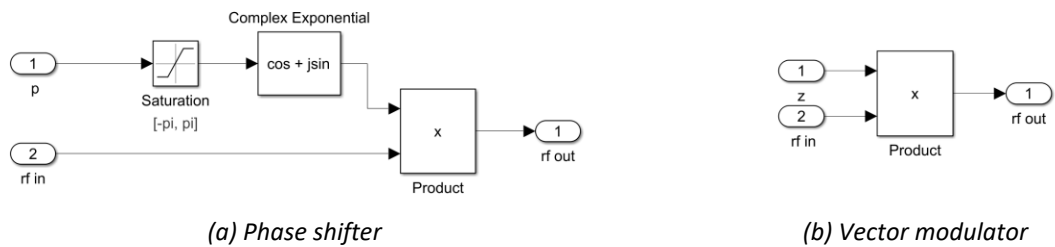

*(a) Phase shifter*          *(b) Vector modulator*

*Figure 2 Simulink envelope model of: (a) a phase shifter with limited range $[-\pi, \pi]$ with a real scalar control variable $p$; (b) a vector modulator with a complex scalar control variable $z = x + iy$.*

Time delay systems are notoriously difficult to control, and TDOs are susceptible to modulation instability. Nevertheless, with careful design informed by Laplace transform analysis of linearized phase models, stable PLL-OEO systems are achievable that meet other desirable performance requirements such as an optimal phase noise spectral density profile. The reader is directed to reference [11] for the analysis and design of stable PLL-OEO systems. The objective herein is to demonstrate continuous tuning without mode hoping of TDOs and the stabilisation of TDOs subject to thermally induced delay drift for which purpose envelope models are more compelling.

---

[5] The solution $\rho = 0$ of Equation 15 is unstable.
[6] That is a locking range limited only by the passband of the RF bandpass filter.

An envelope model represents an analytic signal by the product of a complex envelope and a pure carrier with specified nominal frequency. Since the pure carrier is known completely, it conveys no information. All information on the amplitude and phase fluctuations of the oscillation are encoded by the complex envelope including any detuning from the nominal carrier frequency. Consequently, it is only necessary to model the induced mapping between incoming and outgoing complex envelopes to fully describe a component. The envelope representation is motivated by the sample rate required to avoid aliasing in discrete time simulations. The complex envelope requires a sample rate at least equal to the extent in the frequency domain where the spectral density is significant, which for a $10\ GHz$ OEO is of the order of $2\ MHz$ at most, whereas to adequately represent the same analytical signal with an explicit carrier requires a sample rate of the order of $20\ GHz$.

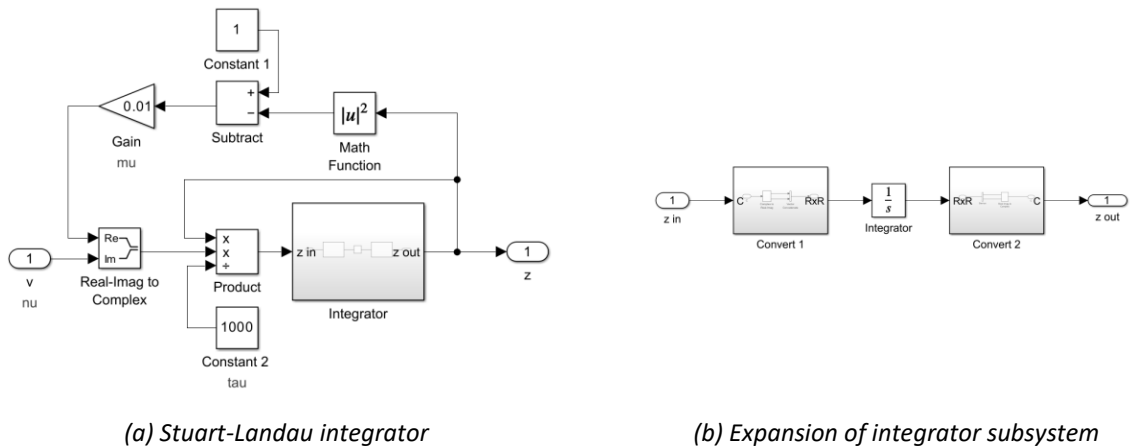

*(a) Stuart-Landau integrator*  *(b) Expansion of integrator subsystem*

*Figure 3 (a) Simulink model of a Stuart-Landau integrator using complex arithmetic; (b) Expansion of the Integrator subsystem block in Figure 3 (a). The Simulink Integrator block is placed between a $\mathbb{C} \rightarrow \mathbb{R}^2$ and a $\mathbb{R}^2 \rightarrow \mathbb{C}$ conversion block as it supports multidimensional real vector but not complex signals.*

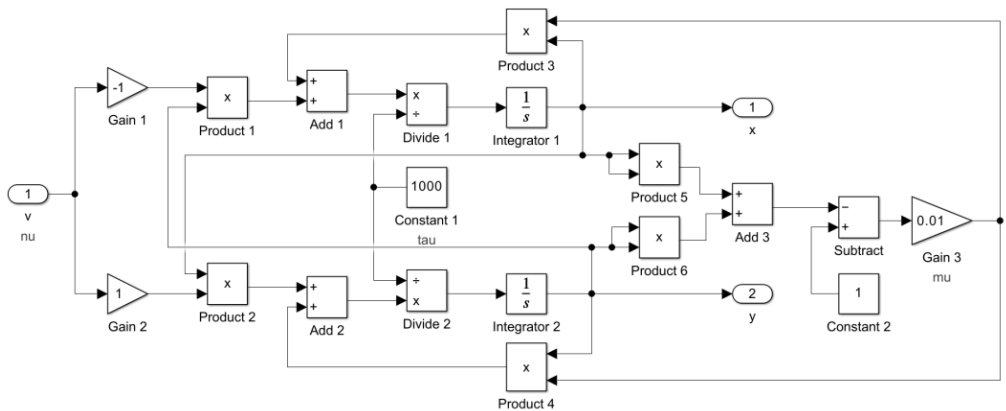

*Figure 4 Simulink model of a Stuart-Landau integrator equivalent to Figure 3 but using real arithmetic. The circuit architecture is consequently amenable to analog electronic implementation using operational amplifier and four-quadrant multiplier integrated circuits.*

The complex envelope representation is used with advantage in optical circuit simulators to avoid having to sample an optical carrier which has a frequency of the order of $200\ THz$. It is tempting to use an optical circuit simulator to simulate an optoelectronic oscillator to take advantage of the sophisticated component models of lasers, modulators, optical fibres, photodetectors, and a more limited range of electronic components. However, the number of samples required to adequately represent the RF modulation in transit through an optical fibre of length of the order of $10\ km$ is prodigious ($\sim 1,000,000$) and simulations of oscillators with fibre lengths of up to $\sim 100\ m$ only are possible before exhausting available computing resources. This problem is avoided by treating the RF-photonic link as an RF delay line; indeed, the purpose of the photonics is to transport the RF carrier over a long path taking advantage of the low loss of the optical fibre. Ideally, the photonics could otherwise be ignored. However, the photonics does contribute to the RF phase noise spectral density through a variety of mechanisms [12].

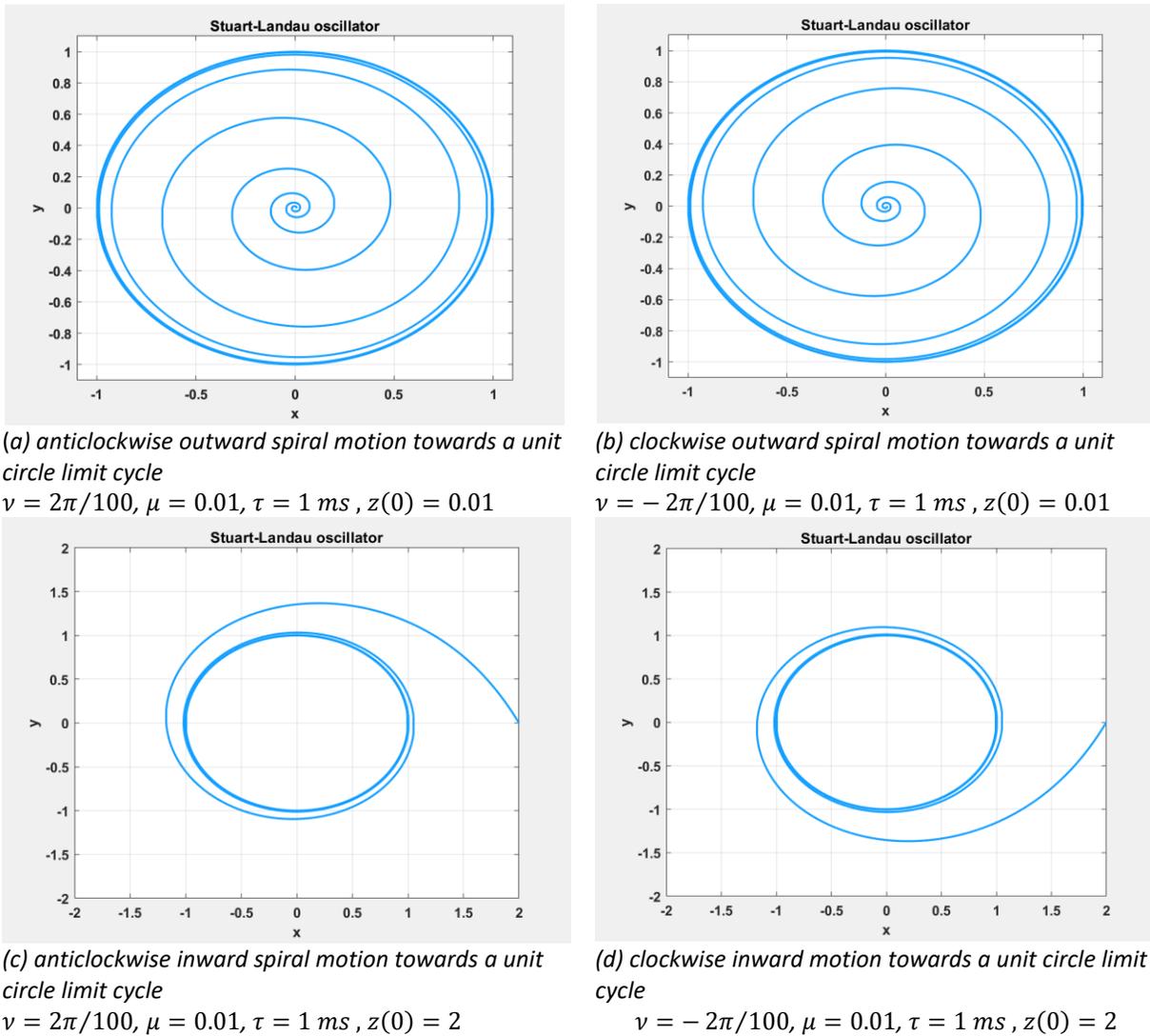

(a) anticlockwise outward spiral motion towards a unit circle limit cycle
$\nu = 2\pi/100, \mu = 0.01, \tau = 1\ ms, z(0) = 0.01$

(b) clockwise outward spiral motion towards a unit circle limit cycle
$\nu = -2\pi/100, \mu = 0.01, \tau = 1\ ms, z(0) = 0.01$

(c) anticlockwise inward spiral motion towards a unit circle limit cycle
$\nu = 2\pi/100, \mu = 0.01, \tau = 1\ ms, z(0) = 2$

(d) clockwise inward motion towards a unit circle limit cycle
$\nu = -2\pi/100, \mu = 0.01, \tau = 1\ ms, z(0) = 2$

*Figure 5 Stuart-Landau oscillator state variable trajectory $z(t)$ on the complex plane illustrating the encirclement of the origin in either direction any number of times and its evolution towards the unit circle.*

Simulink models have been developed of the component parts of a phased locked loop (PLL) containing a phase controlled TDO such the OEO. Figure 1 provides a schematic diagram of a PLL-OEO which features a tuning phase shifter (PS). The Simulink simulations performed involve either a conventional tuning phase shifter with limited range (see Figure 2(a)), a vector modulator configured as a tuning phase shifter (see Figure 2(b)), or both types combined in series.

Continuous tuning & long-term frequency stabilisation of time delay oscillators

Figure 3 provides a block diagram of a Simulink complex-variable model of the Stuart-Landau integrator while Figure 4 provides a block diagram of an entirely equivalent real-variable model. The latter may be reinterpreted as an analog electronic circuit architecture in which standard operational amplifier and four quadrant multiplier subcircuits are used to implement the gain, addition, integration, and product functions. Figure 5 presents the results of simulations confirming the behaviour of the Stuart-Landau integrator as a baseband quadrature phase oscillator with a limit cycle on the unit circle and an instantaneous frequency set by the real control ν. The quadrature components $(x, y)$ distinguish positive frequency ($\nu > 0$ & anti-clockwise rotation) and negative frequency ($\nu < 0$ & clockwise rotation) oscillation.

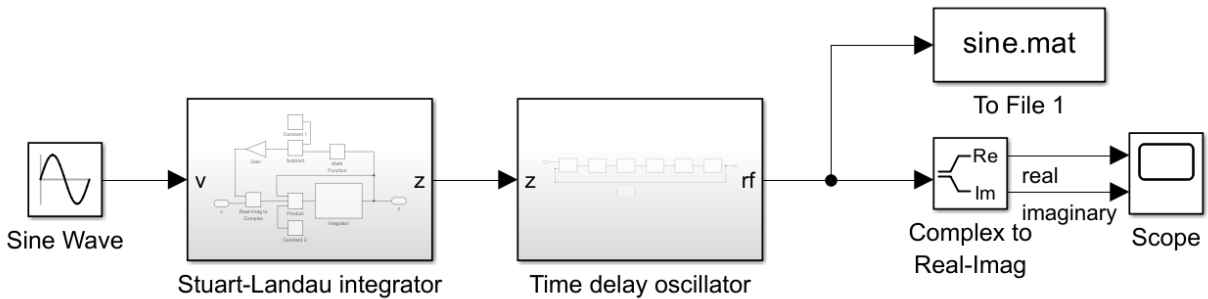

*Figure 6 A Simulink model of a time delay oscillator tuned by an internal vector modulator driven externally by a Stuart-Landau integrator itself driven by a signal generator such that $\nu = a\,cos(\omega t)$ where $a = \pi^2/10$, $\omega = 2\pi\ rad/s$, and t is simulation time. The Stuart-Landau integrator has a time constant $\tau = 1000\ \mu s$ and the time delay oscillator has a round trip group delay $\tau_G = 25\ \mu s$ of which the RF resonator contributes $\tau_R = 0.1\ \mu s$. Consequently, the oscillator supports a comb of 80 resonances with an $FSR = 40\ kHz$ spanning a $10/\pi\ MHz$ passband.*

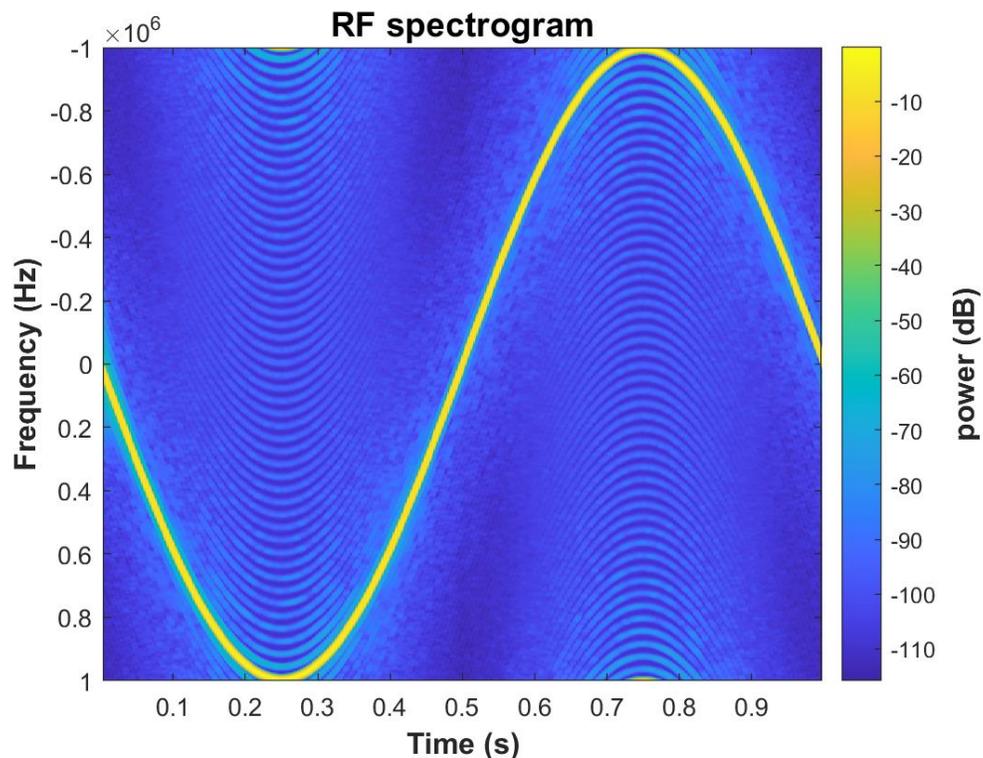

*Figure 7 Spectrogram of the data generated by the Simulink model shown in Figure 6. The main oscillator mode follows a sinusoidal path in the time-frequency domain that spans $\pm 1\ MHz$ ($\pm 25\ FSR$) in frequency with a period of $1s$. The visible sidemode structure is due in part to power law phase fluctuations included in the model of the time delay oscillator and in part to the aggressive rate of change of frequency.*

Figure 6 provides a schematic of a Simulink envelope model of a swept frequency TDO. A sine wave is applied to a Stuart-Landau integrator that tunes the TDO via a vector modulator incorporated within. Figure 7 provides a spectrogram (short-time Fourier transform) of the time domain data generated by the Simulink model. The result demonstrates that a TDO may be tuned over an arbitrary continuous path spanning multiple FSR within the passband of the RF filter by a vector modulator driven by a Stuart-Landau integrator; in this example a sinusoidal path extending over 2 MHz or 50 FSR within a 3 MHz passband.

For comparison both a vector modulator tuned TDO under vector PLL control, and a conventional phase-shifter tuned TDO under scalar PLL control are modelled. Figure 8 and Figure 9 provide schematics of their respective Simulink envelope models. The parameters of both PLLs are identical (see Table 1).

**Time delay oscillator**

| | |
|---|---|
| $\tau_D = 24.9 \ \mu s$ | Delay line delay time |
| $\tau_R = 0.1 \ \mu s$ | Resonator on-resonance group delay |
| $\Delta f = 1/\pi\tau_R = 3.18 \ MHz$ | Resonator bandwidth |
| $\tau_G = 25 \ \mu s$ | Round trip group delay time |
| $FSR = 1/\tau_G = 40 \ kHz$ | Frequency interval between modes |
| $\partial f/\partial t = 1 \ MHz/s$ | Thermally induced delay drift induced natural frequency chirp |

**Phase locked loop**

| | |
|---|---|
| $N = 100$ | Frequency division ratio |
| $\tau_I = N \ ms$ | Integration time constant |
| $\kappa = \sqrt{5}/N$ | Proportional gain |
| $\tau_F = \sqrt{\tau_G\tau_I} = \sqrt{2.5} \ ms$ | System time constant |
| $f_n = 1/2\pi\tau_F = 100.7 \ Hz$ | System natural frequency |
| $\xi = (1/2)\kappa\sqrt{\tau_I/\tau_F} = 1/\sqrt{2}$ | System damping factor |
| $\tau_{3,4,5} = 80, 40, 20 \ ms$ | Time constants associated with three additional system poles ($s\tau_{3,4,5} = 1$) introduced by the low-pass filter |

*Table 1 Phase locked loop model parameters*

There is a subtlety in the description of delay drift associated with the translation to zero frequency of the nominal frequency of the oscillator. In practice, the delay is large on the scale of the period of the oscillation. Consequently, a small relative variation of the delay can be significant relative to the oscillation period. In the case of a $5 \ km$ optical fibre, a $0.1\%$ variation of length corresponds to a $50 \ cm$ length change that results in a $50\pi$ radian phase change of a $10 \ GHz$ RF carrier, yet the $40 \ kHz$ FSR changes by only $40 \ Hz$. With frequency transposed to baseband, a $0.1\%$ change in the $25 \ \mu s$ delay will have negligible effect on the simulation. Thermally induced delay drift can only be simulated through the phase shift it introduces and not by varying the delay time within the simulation. Consequently, use of an indefinite range phase shifter (Figure 2 with the *Saturation* block removed) driven by a ramp source within the TDO is used to simulate delay drift in the PLL simulations.

The controller provides proportional integral control to the TDO to attempt to counter the thermally induced delay drift. The phase detector within the controller generates an unwrapped phase difference that is divided by $N$ before being limited to a range of $[-2\pi, 2\pi]$. This emulates a sequential phase detector with a frequency divider in the feedback loop. Since the divider has been moved to the output of the phase detector, an implicit $\times N$ frequency multiplier is present in the reference path. The phase fluctuations of the OCXO are consequently set to $\times N$ of those reported in reference [10]. The presence

Continuous tuning & long-term frequency stabilisation of time delay oscillators of the divider in the feedback loop results in the proportional gain being equal to the set value $\times\ 1/N$ and the integration time constant being equal to the set value $\times N$.

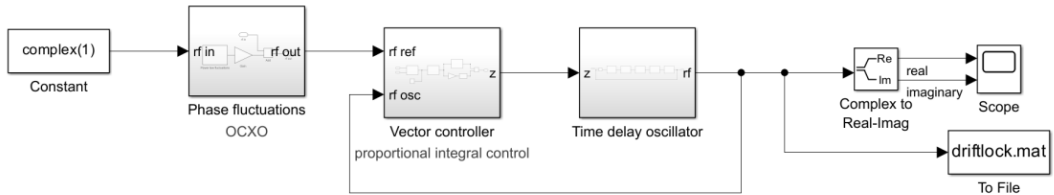

(a) Complete system

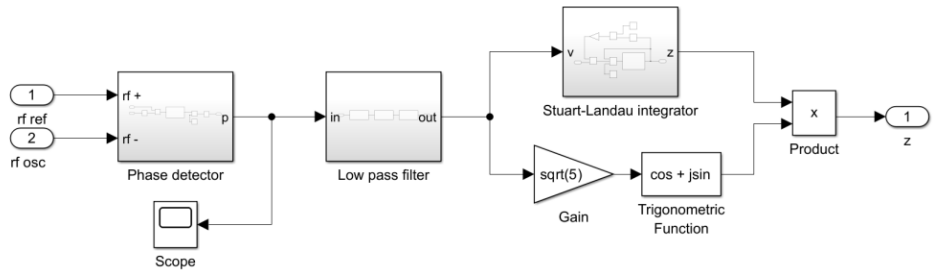

(b) Expansion of the vector controller subsystem

Figure 8  Block diagram of a Simulink model of a vector modulator tuned time delay oscillator under vector PLL control.

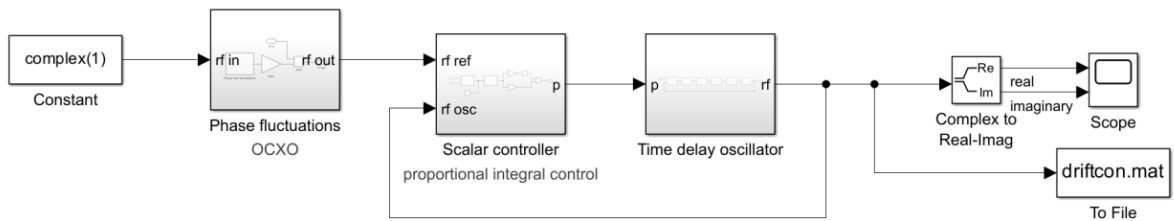

(a) Complete system

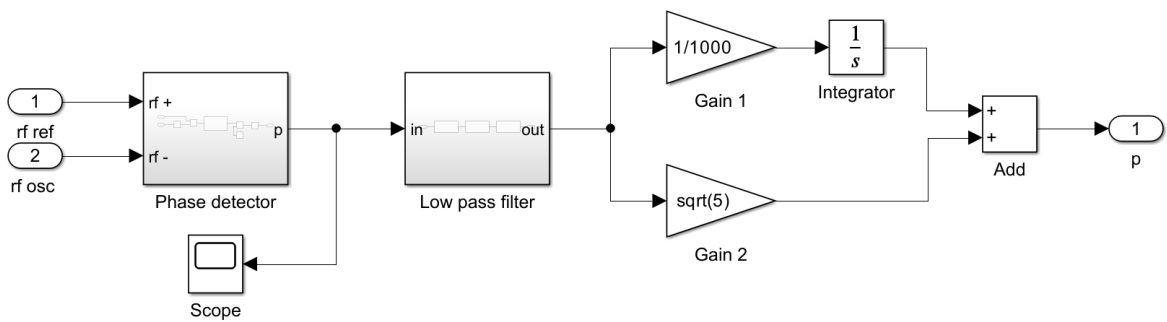

(b) Expansion of the scalar controller subsystem

Figure 9 Block diagram of a Simulink model of a conventional phase shifter tuned time delay oscillator under scalar PLL control.

Figure 10 provides a comparison of the time domain evolution of the scalar and vector PLL systems. Both systems achieve lock after 15 $ms$ but the conventional phase shifter tuning & scalar controller loses lock after 20 $ms$ when it reaches the limit of its phase-shifter's range $[-\pi, \pi]$ whereas the vector modulator tuning & vector controller maintains solid lock indefinitely. Figure 11 provides spectrograms of the frequency drifting free oscillator and vector controller locked oscillator over a more prolonged time interval (1 $s$). Figure 11(a) shows that the natural frequency of the oscillator is drifting at a rate of $1 MHz/s$. The limited range of the phase-shifter can only compensate drift over one FSR $[-20\ kHz, 20\ kHz]$ the limits of which are reached after 20 $ms$ consistent with the loss of lock observed in Figure 10(a). The vector modulator in contrast continues its trajectory around the unit circle and hence maintains lock indefinitely as observed in Figure 10(b) and Figure 11(b).

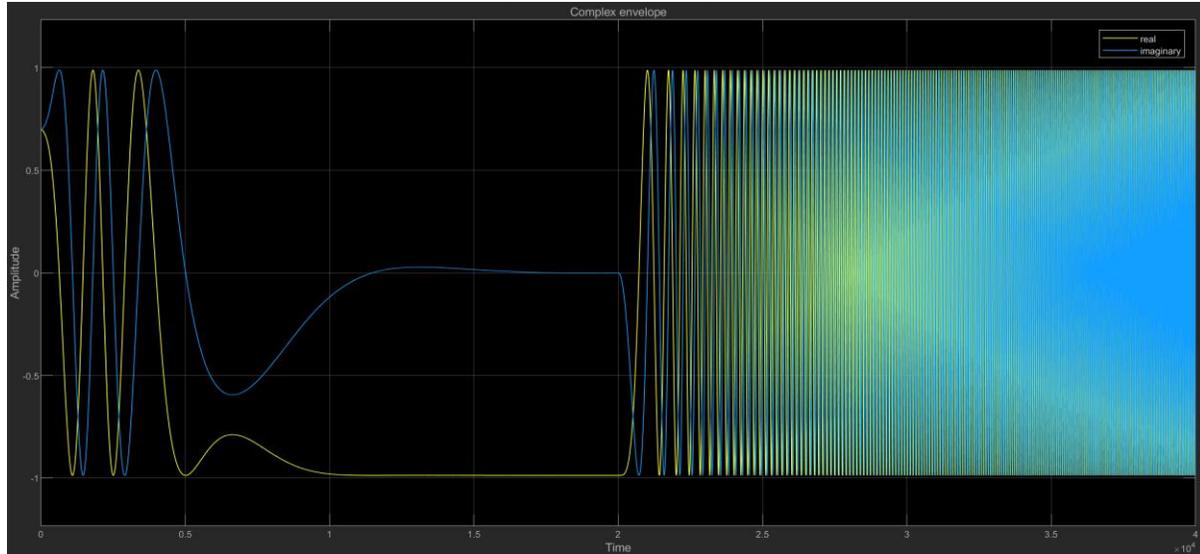
(a) conventional phase shifter tuning and scalar controller

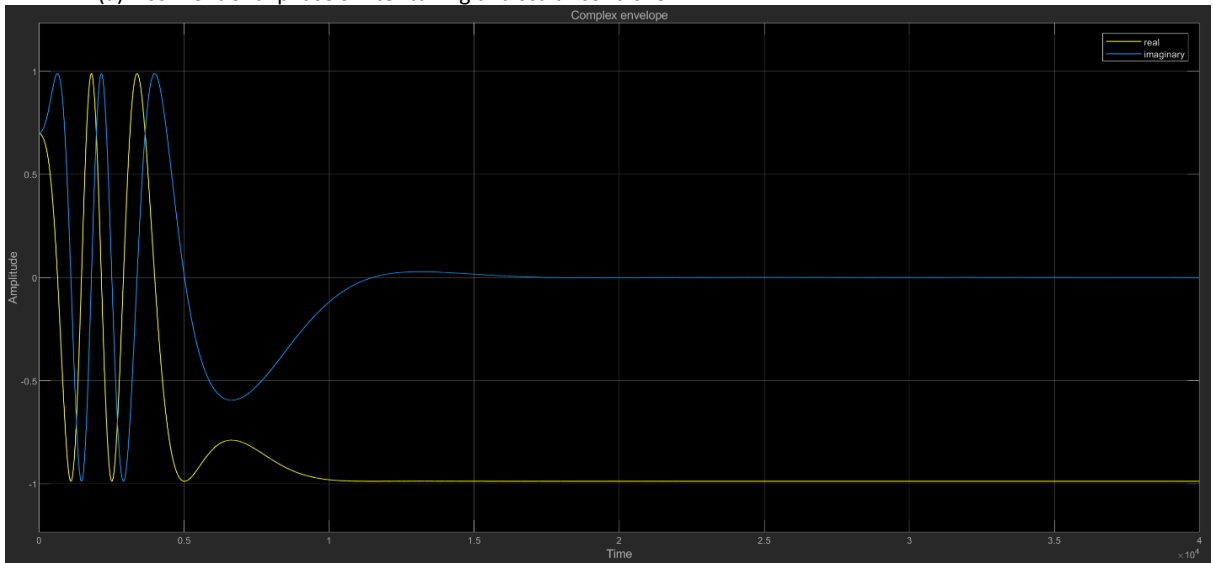
(b) vector modulator tuning & vector controller

*Figure 10 Evolution of the complex envelope the output of a time delay oscillator under PLL control. The thermally induced delay drift results in a $1\ MHz/s$ rate of change of the natural frequency of the free oscillator.*

Figure 12 provides the phase noise spectrum of the vector modulator tuned TDO under PLL control. Power law phase fluctuations are implemented to reproduce experimentally measured phase noise spectral densities of both the multiplied reference source and the TDO where they are applied within the oscillator. The phase noise at frequencies below 200 Hz is suppressed by the action of the PLL. Otherwise, the phase noise spectrum is that of a free TDO in the absence of thermally induced delay drift, as is desired.

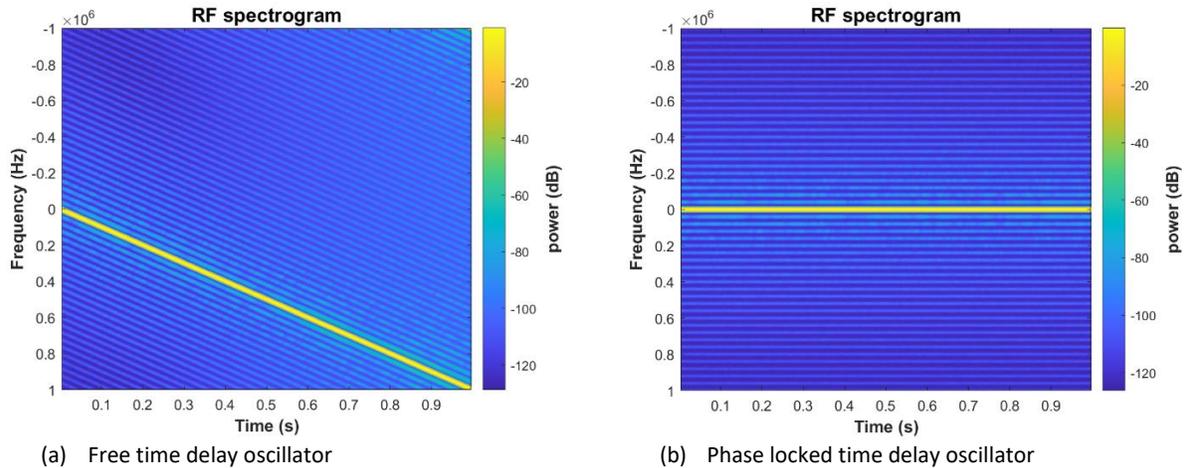

(a) Free time delay oscillator  (b) Phase locked time delay oscillator

*Figure 11 Spectrogram of the RF output of a time delay subject to thermally induced delay drift. (a) The thermally induced delay drift induces a $1\ MHz/s$ rate of change of the natural frequency of the free oscillator. (b) The vector modulator tuning & vector controller overcome the induced natural frequency drift and maintain solid lock indefinitely.*

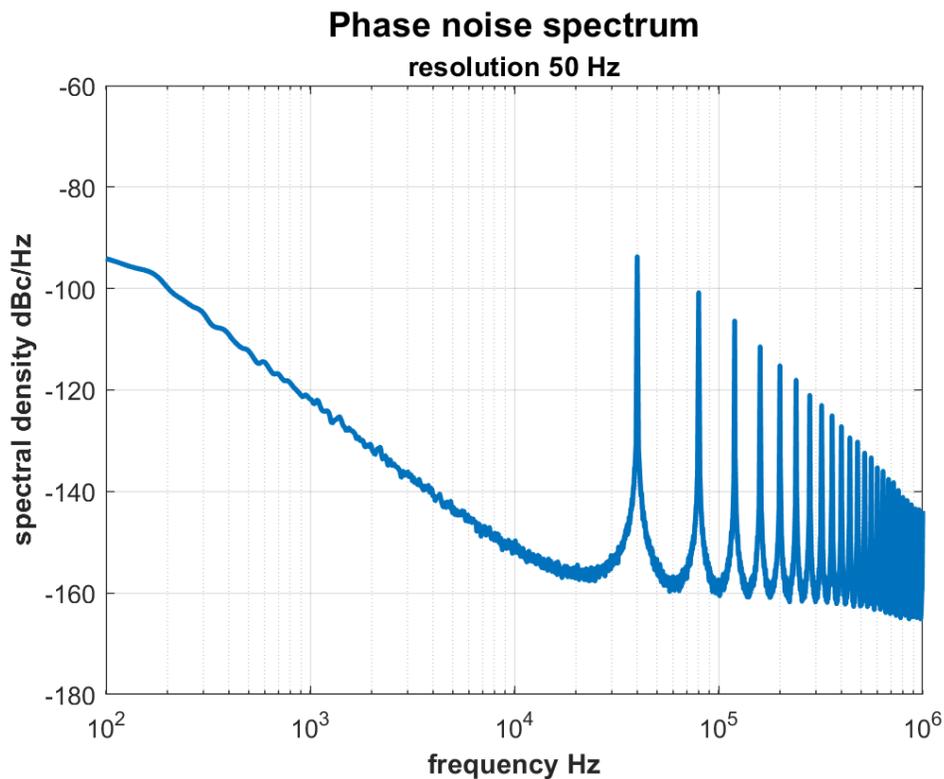

*Figure 12 Phase noise spectral density of the vector modulator tuned time delay oscillator under vector PLL control.*

The drift implemented in the simulations is necessarily aggressive to avoid excessively prolonged simulation execution times. Nevertheless, the same process leading to loss of lock due to a bounded phase shifter range plays out in practice on a longer time scale. Experimental trials conducted at Nanowave Technologies Inc. found that a prototype phase shifter tuned OEO controlled by a conventional PLL lost lock when subject to a 5°C increase in ambient temperature. In contrast a vector modulator tuned version of the same OEO maintained solid lock even when placed in an oven and cycled over a temperature range from ambient to 80 °C. A patent application has been filed [13] on this technique which enables practical deployment of the OEO technology into high performance RADAR and communication systems without recourse to complex high power consumption temperature controlled enclosures.

## Conclusions

The phase shifter used conventionally to tune an OEO has a finite range that is insufficient to compensate thermal drift of the delay of a long optical fibre over the operational temperature range. Without a temperature-controlled enclosure, such an OEO placed within PLL will lose lock within minutes. A solution is proposed in this paper that provides an indefinite extension to the tuning range. In essence the polar co-ordinate system (magnitude $\rho = 1$ & phase $\theta$) corresponding to the phase shifter is replaced by a Cartesian co-ordinate system (in-phase $x$ & quadrature-phase $y$) corresponding to a vector modulator with $(x, y)$ constrained to lie on the unit circle. The trajectory defined by the vector modulator may wind any number of times either way around the origin while all physical variables remain bounded enabling continuous tuning without mode hops and long-term stabilisation without temperature control. The concept has been fully verified by Simulink simulations and proven experimentally using a prototype OEO which maintained lock to a system reference indefinitely over a wide temperature range. The solution is paradigm shifting because it renders practical new strategies for tuning complex time delay oscillator systems in general and specifically optoelectronic oscillators.

## Acknowledgements

Mehedi Hasan and Trevor J. Hall are indebted to Nanowave Technologies Inc. for providing access to their prototype optoelectronic oscillator to test the tuning solution presented in this paper. Mehedi Hasan is grateful to the Natural Sciences and Engineering Research Council of Canada (NSERC) for their support through the Vanier Canada Graduate Scholarship program. Trevor J. Hall is grateful to the University of Ottawa for their support of a University Research Chair.